\documentclass[9pt,conference]{IEEEtran}
\usepackage{waspaa25}

\usepackage{bm} % for bold math symbols (incl. Greek letters) with \bm{}

\title{The Perception of Phase Intercept Distortion and its Application in Data Augmentation}

\name{Venkatakrishnan Vaidyanathapuram Krishnan,
      Nathaniel Condit-Schultz
      }
\address{Georgia Institute of Technology, Atlanta, USA}

\begin{document}

\maketitle

\begin{abstract}
% To help authors prepare their manuscripts for submission to WASPAA 2025 and generate IEEE Xplore-compatible PDFs, we compile a list of guidelines and put together two templates for users of both \LaTeX\ and Microsoft Word, which can be downloaded from the workshop website at \cite{waspaaweb}. These guidelines and templates are modified from those for ICASSP and past WASPAA workshops, so an experienced author who has published something in these conferences/workshops will find it easy to follow the guidelines and use the templates. Note that the WASPAA 2025 style file is now based on IEEEtran.cls, and that there have been many changes to the formatting this year. Please make sure to adhere to the new formatting style.
Phase distortion refers to the alteration of the phase relationships between frequencies in a signal, which can be perceptible.
In this paper, we discuss a special case of phase distortion known as phase-intercept distortion, which is created by a frequency-independent phase shift.
We hypothesize that, though this form of distortion changes a signal's waveform significantly, the distortion is imperceptible.
Human-subject experiment results are reported which are consistent with this hypothesis.
Furthermore, we discuss how the imperceptibility of phase-intercept distortion can be useful for machine learning, specifically for data augmentation.
We conducted multiple experiments using phase-intercept distortion as a novel approach to data augmentation, and obtained improved results for audio machine learning tasks.

% Acknowledgements: Mention if AI was used for grammar enhancement

\end{abstract}

\section{Introduction}
\label{sec:introduction}

In general, any unintentional distortion of signals in audio technology is undesirable.
However, the human auditory system is not sensitive to all forms of distortion and this insensitivity can be leveraged in the engineering of human-facing audio systems.
Here, we consider the case of \emph{phase distortion} \cite{ohm1843definition} which occurs when the phase response of a system is nonlinear \cite{nyquist1930measurement-phase-distortion}, distorting the phase relationships between frequency components in a signal \cite{preis1982phase-distortion-tutorial}. 
A distortionless system's discrete time impulse response can be defined as:
\begin{equation}
    h[n] = K\delta[n-\tau]
    \label{eq:distortionless_system}
\end{equation}
Here, $\delta[n]$ is the impulse function, $K>0$ is the gain parameter corresponding to a constant magnitude response, and $\tau \geq 0$ is the time-delay constant corresponding to a linear phase response with a slope of $-\tau$. 
If the system's phase response is not linear, phase distortion occurs, which can lead to noticeable changes in perceived timbre \cite{plomp1969perception_complex_tone_phase, deer1985perception-phase-distortion-all-pass, preis1983perception-phase-distortion-anti-alias, lipshitz1982audibility-phase-intercept-distortion, craig1962perception_two_tone_phase, hansen1974aural}.
However, it has long been observed that phase-distortion can be imperceptible in some cases \cite{von1912helmholtz, deer1985perception-phase-distortion-all-pass}.

% Distortion refers to the alteration of the original waveform of an audio signal.
% Phase distortion is the degradation of a waveform by altering the phase relationships between the frequency components \cite{preis1982phase-distortion-tutorial}.
% In mathematical terms, phase distortion occurs when the phase response of a system is not linear \cite{nyquist1930measurement-phase-distortion}.

Though human hearing is binaural, and inter-aural phase differences are crucial for spatial awareness and sound localization \cite{bregman1994auditory-scene-analysis}, we limit our discussion to monaural phase effects, where identical signals arrive at both ears.
Monaural perceptual studies have shown that altering the relative phase between sinusoids affects perceived timbre---whether in simple two-tone signals \cite{craig1962perception_two_tone_phase, lipshitz1982audibility-phase-intercept-distortion} or in the harmonics of complex sounds \cite{plomp1969perception_complex_tone_phase}.
The monaural perceptibility of phase distortion in all-pass filters \cite{deer1985perception-phase-distortion-all-pass, suzuki1980perception-phase-distortion}, anti-alias filters \cite{preis1983perception-phase-distortion-anti-alias}, and speech enhancement systems \cite{chappel2016phase-distortion-speech} has also been studied.
% Clearly, nonlinearly modifying the phase relationships between sinusoids can cause changes in perceived timbre.
However, distortion caused by a constant phase shift across all frequencies has not been extensively studied.
In this paper, we present evidence that the special case of \emph{phase-intercept distortion} is not perceptible in real-world sounds, and show how this fact can be leveraged for data augmentation in audio-based machine learning applications.

\subsection{Phase-intercept Distortion}

For a single sinusoidal tone, a constant phase shift is defined as:

\begin{equation}
    x(t) = \sin(\omega_0 t + \phi) = \frac{1}{2i}(e^{i\omega_0 t}e^{i\phi} - e^{-i\omega_0 t}e^{-i\phi})
\end{equation}

Note that we must add the positive phase to the positive frequencies and the negative of that phase to the negative frequencies.
This can be verified by using a single sine tone and the effect of a shift in phase in its frequency domain.
\begin{equation}
    \hat{x}(\omega) = \frac{1}{2i}[\delta(\omega - \omega_0)e^{i\phi} - \delta(\omega + \omega_0)e^{-i\phi}]
\end{equation}
This operation is called the frequency-independent phase shift, and can be performed using the \textit{signum} function, defined as:
\begin{equation}
    sgn(\omega) = \begin{cases}
        1 & ;\omega > 0 \\
        0 & ;\omega = 0 \\
        -1 & ;\omega < 0
    \end{cases}
    \label{eq:signum_function}
\end{equation}
The transfer function of a frequency-independent phase shift of $\theta$ can then be defined as:

\begin{equation}
    |H(\omega)| = 1; \hspace{10pt}\Phi(\omega) = \theta \cdot sgn(\omega) 
    \label{eq:freq-response-phase-intercept-distortion}
\end{equation}
The phase response of this operation is piecewise constant and is nonlinear.
Thus, frequency-independent phase shifting creates distortion and is called \emph{phase-intercept distortion}, a special case of phase distortion.
This operation results in a group delay of zero, but a nonlinear phase delay for all non-zero frequencies.
%Although this operation has a group delay of zero, they do not satisfy the distortionless condition shown in \cref{eq:distortionless_system} and are thus known to ``distort'' a signal.

Let the Fourier transform be denoted as $\mathcal{F}$.
Applying this operation on an input signal $x(t)$ with its Fourier Transform $\mathcal{F}(x) = \hat{x}(\omega)$ will result in a rotated signal:
\begin{equation}
    x_{\theta}(t) = \mathcal{F}^{-1}(\hat{x}(\omega)e^{i \theta \cdot sgn(\omega)})
    \label{eq:freq-ind-phase-shift}
\end{equation}
% When we input \cref{eq:freq-response-phase-intercept-distortion} into the group delay and phase delay equations, we get:
% \begin{equation}
%     \tau_p = - \frac{\Phi(\omega)}{\omega} = - \frac{\theta}{\omega}
% \end{equation}
% \begin{equation}
%     \tau_g = - \frac{d\Phi(\omega)}{d\omega} = - \frac{d(\theta)}{d\omega} = 0
% \end{equation}
A popular example of a frequency-independent phase-shift operation is the Hilbert Transform.
The Hilbert transform introduces a phase shift of $-90^\circ$ for the positive frequencies and $90^\circ$ for the negative frequencies, without altering the signal's magnitude spectrum.
Let $\mathcal{H}$ denote the Hilbert transform and consider a signal $x(t)$ with its Fourier transform $\hat{x}(\omega)$; the Hilbert Transform modifies the complex spectral content as follows:
\begin{equation}
    \mathcal{F}(\mathcal{H}(x)) = -i\hat{x}(\omega) \cdot sgn(\omega)
    % \begin{cases}
    %     -i\hat{x}(\omega) & ; &\omega>0 \\
    %     0 & ; &\omega=0 \\
    %     i\hat{x}(\omega) & ; &\omega<0 \\
    % \end{cases}
    \label{eq:hilbert_transform}
\end{equation}
% where $sgn(\omega)$ is the \textit{signum} function as defined in \cref{eq:signum_function}.

A useful property of the Hilbert transform of a real-valued signal $x(t): \mathbb{R} \rightarrow \mathbb{\mathbb{R}}$, is that it is orthogonal to the original signal, as can be easily proved by showing that its inner product with $\mathcal{H}(x(t))$ is equal to zero.
% Consider a real-valued signal $x(t): \mathbb{R} \rightarrow \mathbb{\mathbb{R}}$.
% To demonstrate orthogonality, we must prove its inner product with $\mathcal{H}(x)(t)$ is equal to zero:
% \begin{equation*}
%     \langle x, \mathcal{H}(x) \rangle 
%     = \langle \mathcal{F}(x), \mathcal{F}(\mathcal{H}(x)) \rangle 
% \end{equation*}
% \begin{equation*}
%     = \int_{-\infty}^0 \hat{x}(\omega) \overline{i\hat{x}(\omega)} d\omega + \int_{0}^\infty \hat{x}(\omega) \overline{-i\hat{x}(\omega)} d\omega
% \end{equation*}
% \begin{equation*}
%     = -i \int_{-\infty}^0 \hat{x}(\omega) \overline{\hat{x}(\omega)} d\omega + i\int_{0}^\infty \hat{x}(\omega) \overline{\hat{x}(\omega)} d\omega
% \end{equation*}
% \begin{equation*}
%     = i \bigg(\int_{0}^\infty |\hat{x}(\omega)|^2  d\omega - \int_{-\infty}^0 |\hat{x}(\omega)|^2  d\omega \bigg)
% \end{equation*}
% Since $x(t)$ is real-valued, the magnitude spectrum $|\hat{x}(\omega)|$ must be an even function.
% \begin{equation*}
%     \langle x, \mathcal{H}(x) \rangle = i \bigg(\int_{0}^\infty |\hat{x}(\omega)|^2  d\omega - \int_{0}^\infty |\hat{x}(\omega)|^2  d\omega \bigg) = 0
% \end{equation*}
% Hence proved that the Hilbert transform must be orthogonal to any real-valued signal.
This property makes it fundamental in applications such as analytic signal representation \cite{bedrosian2007analyticrespresentation, gabor1946theoryofcommunication}, envelope detection \cite{yang2017analyticsignalenvelope}, phase retrieval \cite{taylor1981phase}, single-sideband (SSB) modulation, and instantaneous frequency estimation \cite{huang2009instantaneousFrequencyAnalytic}.

We can use the Hilbert transform to create an analytic signal representation of the original signal \cite{gabor1946theoryofcommunication} as follows:
\begin{equation}
    x_a(t) = x(t) + i \mathcal{H}(x(t))
    \label{eq:analytic_signal_representation}
\end{equation}
The frequency-independent phase shift operation can be achieved by directly applying a rotation of $\theta$ to the analytic signal and taking its real part:
\begin{equation}
    x_{\theta}(t) = Re[x_a(t) e^{i \theta}]
    \label{eq:freq_ind_phase_shift}
\end{equation}
This can be shown by expanding \cref{eq:freq-ind-phase-shift} and rearranging the terms.
% \begin{equation*}
%     \hat{x_{\theta}}(\omega) = \hat{x}(\omega)e^{i \theta \cdot sgn(\omega)}
% \end{equation*}
% \begin{equation*}
%     = \hat{x}(\omega)[\cos (\theta \cdot sgn(\omega)) + i \sin (\theta \cdot sgn(\omega))]
% \end{equation*}
% \begin{equation*}
%     = \begin{cases}
%         \hat{x}(\omega)\cos \theta + i \hat{x}(\omega) \sin \theta & ; \omega > 0 \\
%         \hat{x}(0)\cos \theta & ; \omega > 0 \\
%         \hat{x}(\omega)\cos (\theta) - i \hat{x}(\omega) \sin \theta & ; \omega > 0 \\
%     \end{cases}
% \end{equation*}
% Using \cref{eq:hilbert_transform} and taking the inverse Fourier transform, we obtain:
% \begin{equation*}
%     x_{\theta}(t) = x(t)\cos \theta - \mathcal{H}(x)(t) \sin \theta
% \end{equation*}
% \begin{equation*}
%     = Re[x_a(t) e^{i \theta}]
% \end{equation*}
In summary, phase-intercept distortion can be modeled by constructing the analytic representation of the signal using the Hilbert transform, rotating it, and extracting the real part of this complex signal \footnote{https://github.com/Computational-Cognitive-Musicology-Lab/Phase-Intercept-Distortion}.

\subsection{Perception of Phase-intercept Distortion}

% The experiments discussed previously were focused on systems that perform frequency-dependent phase shifts.
% We know that phase-intercept distortion arises in the context of a frequency-independent phase shift.
The perception of phase-intercept distortion has not been explored as much as other common forms of phase distortion.
One specific case that has been studied is polarity reversal---corresponding to a phase shift where $\theta = \pm 180^{\circ}$ in \cref{eq:freq_ind_phase_shift}.
Though most researchers (and audio engineers) assume that polarity reversal is inaudible, which is consistent with experimental evidence \cite{sakaguchi2000polarity-inversion-speech-perception}, there have also been some claims that it is audible \cite{lipshitz1982audibility-phase-intercept-distortion, craig1962perception_two_tone_phase}.
The Wood Effect\footnote{Named after Charles L. Wood.} \cite{craig1962perception_two_tone_phase} is one dramatic edge case where polarity reversal creates a minute difference in timbre for lower frequencies.
This effect occurs when inverting a sine wave which has been clipped on only one half of each cycle;
The resulting polarity-inverted signals sound slightly different, showing that our ears do treat the clipped portion differently when it is presented as a compression or as a rarefaction.
No other cases of perceivable phase inversion have been reported.

% Studies on perception of polarity reversal
%The perception of phase-polarity reversal in an audio signal has been studied to a limited extent.
%Polarity reversal, being a special case of phase-intercept distortion, is still considered by many to be a form of distortion.
%An experiment was conducted using an algorithm to detect syllable boundaries and randomly invert or retain each syllable signal without introducing discontinuities.
%The human participants could not distinguish between the original and polarity-inverted speech signals \cite{sakaguchi2000polarity-inversion-speech-perception}.

\begin{figure}
    \centering
    \includegraphics[scale=0.15]{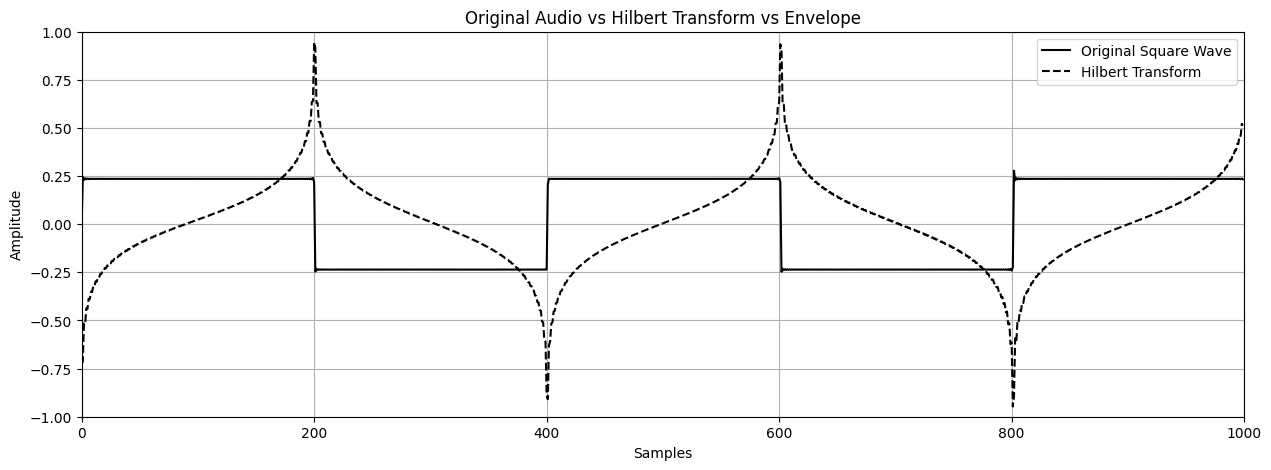}
    \caption{Illustration of the effect of the Hilbert Transform ($\theta = \pm 90^{\circ}$) on a square wave.}
    \label{fig:square_wave_hilbert}
\end{figure}

Phase-intercept distortion can have a large impact on an audio waveform (\cref{fig:square_wave_hilbert}).
However, human hearing relies on both time-domain and frequency-domain characteristics of sound \cite{lipshitz1982audibility-phase-intercept-distortion}, and visible alterations to the waveform shape may not necessarily correlate with changes to the perceived sound.
To our knowledge, the Wood Effect is the only case of perceptible phase-intercept distortion which has been reported in the literature.
Research on the perception of frequency-independent phase shifts at arbitrary angles (i.e., besides $\theta = \pm 180^{\circ}$) has not been reported.
In informal testing, we observed that arbitrary-angle phase-intercept distortion does not seem to be perceptible in real-world audio examples.
We thus hypothesized that frequency-independent phase shifting is imperceptible for general audio signals.
In the next section, we report the results of an experiment to test this hypothesis by measuring human participants' ability to detect phase-intercept distortion in ecologically-valid sounds.

% To our knowledge, research on the perception of other forms of frequency-independent phase shifts, such as the Hilbert transform (a constant $-\pi/2$ radians phase shift on all frequencies) and general frequency-independent phase shifts, has not been conducted.

%One can even see the effect of the Hilbert transform performed on a square wave in \cref{fig:square_wave_hilbert}.
% The existing literature on the perceptual effects of different types of phase distortion is somewhat inconsistent, with much of it being quite old, dating back to the 1950s or earlier.
% Since basic assumptions about the perception of phase undergird many areas of audio and music research, the lack of clarity in the literature is problematic.
% For our purposes, lack of clarity about the perception of phase-intercept distortion is particularly problematic.

% 
% (Any paper on saying subjective evaluation must match objective evalution?)
% 

% There have been experiments performed, and reported to show a difference for acoustic scenarios?? \cite{lipshitz1982audibility-phase-intercept-distortion}, but this wouldn't be necessarily true for source separation tasks.
\section{Experiment}
\label{sec:experiment}

% The existing literature on the perceptual effects of different types of phase distortion is somewhat inconsistent.
% Much of the literature is also quite old, dating back to the 1950s or earlier.
% Since basic assumptions about the perception of phase undergird many areas of audio and music research, the lack of clarity in the literature is problematic.
% For our purposes, lack of clarity about the perception of phase-intercept distortion is particularly problematic.
% Based on preliminary informal testing, phase-intercept distortion did not seem to be perceptible in real-world audio examples.
% % contradicting xxx's \cite{hansen1974aural}.
% Based on our hypothesis, we conducted a perceptual experiment to attempt to verify (or refute) this observation.
% Specifically, we hypothesized that phase-intercept distortion is \emph{not} perceptible in general, real-world cases.

\subsection{Participants}

Forty-eight participants were recruited via different professional and university mailing lists. \footnote{This experiment was conducted with the approval of the Georgia Tech's Institution Review Board (IRB) (ethics board).}
Participants were only required to be physically present in the United States, at least 18 years old, and have no history of hearing loss (screened through self-report).
Only 27 of our 48 participants completed 100\% of the survey and we discarded data from one participant who reported a hearing disability, leaving complete data from 26 participants.

%To ensure data completeness and consistency in the results, only data from those 24 participants were analyzed.

We anticipated that individual differences between participants (age, gender, music background, etc.) might affect their ability to distinguish phase-intercept distortion.
Thus, through a pre-questionnaire, we collected information about the participants' age, gender, and experience with music, musical instruments, music production, mixing engineering, and high-fidelity listening habits.
Twenty-two participants identified as male (mean age: 30.4 years) and the rest as female (mean age: 40.8 years); eight of the participants were above the age of 35.
Fourteen participants identified as musicians.
In the final analysis, none of these factors seem to affect participants' performance.

\subsection{Stimuli}
\label{sec:chap3-datasets}

For stimuli, we gathered audio recordings of a variety of music, speech, and general sounds (e.g., traffic or machine sounds) from several existing datasets.
We then manipulated these recordings by applying phase-intercept distortion as needed.
We sampled ten recordings each of music, speech, and real-world sounds, for a total of thirty.
The majority of the samples were taken from AudioSet \cite{gemmeke2017audioset}, one of the largest available collections of music and speech audio samples.

Our ten \textit{music} samples were a mix of monophonic and polyphonic recordings of instruments or a combination of instruments such as guitar, bass, drums, etc.
These recordings were randomly sampled from a popular source separation dataset, the MUSDB18-HQ dataset \cite{rafii2019musdb18hq}---both mixes and isolated stems---with a few more from the AudioSet.
Phase distortion generally ``smears'' the signal \cite{lipshitz1982audibility-phase-intercept-distortion}, so its effect will be most prominent on percussive sounds and transients \cite{mathes1947perception_mpe}.
Thus, three purely percussive recordings were included along with seven music recordings that also contain percussion in the mix.
To further ensure that attacks are not ``smeared'' off, tonal percussion was also included by randomly sampling pure percussion samples from multi-tracks in the Saraga \cite{srinivasamurthy2021saraga} and Sanidha \cite{krishnan2025sanidha} datasets,
 which include clean, isolated recordings of tonal percussion like the \textit{mridangam}.
%These are Indian Music datasets, which contain tonal percussion multitracks such as the \textit{mridangam}, \textit{tabla} and \textit{ghatam}.
% The primary reason for using these datasets, instead of using the ethnic music audio samples directly from AudioSet, is that they provide cleaner and isolated tracks of these tonal percussion instruments.
%The primary reason for using these datasets is that they provide cleaner and isolated tracks of these tonal percussion instruments.

Our ten \textit{speech} samples were drawn from two sources: seven were sampled randomly from the AudioSet and three were sampled from the Librispeech \cite{panayotov2015librispeech} dataset.
We included Librispeech samples beause they are clean and isolated, contrasting with the speech recordings from AudioSet, which include chatter, chants, generic conversations, etc.
The ten \textit{other} samples were randomly chosen from the remaining sounds in AudioSet.

All stimuli recordings used 32-bit depth audio, with sample rates varying depending on the original recording source:
recordings from Librispeech use a sample rate of 16 kHz;
recordings from AudioSet use either a 44.1 kHz or 48 kHz sampling rate; the rest of the recordings use a 44.1 kHz sampling rate.

\subsection{Audio Editing and Manipulation}

To make experimental stimuli short enough for participants to easily hold them, and compare them, in working memory, a three-second excerpt was extracted from each recording.
Many of the recordings have silent moments, so to prevent sampling silence, we repeatedly sampled a random starting point from $[0, l-3]$ (where $l$ is the length of the recording in seconds) until a non-empty three-second audio clip was obtained.
``Non-emptiness'' was defined by the $l_2$-norm of the signal exceeding a chosen threshold value.

To obtain the phase-intercept distorted stimuli, we randomly sampled thirty $\theta$ values from a uniform distribution, ranging from $-\pi$ to $\pi$.
\begin{equation*}
    \theta \sim Unif(-\pi, \pi)
\end{equation*}
We computed the Hilbert transform of each of the thirty stimuli to construct the analytic signals using \cref{eq:analytic_signal_representation}.
To finally apply phase-intercept distortion, the sampled $\theta$ values were then input to \cref{eq:freq_ind_phase_shift}.
% \begin{equation*}
%     x_a(t) = x(t) + i \mathcal{H}(x)(t)
% \end{equation*}
% \begin{equation*}
%     x_{\theta}(t) = Re[x_a(t) e^{i \theta}]
% \end{equation*}

% Fade in and outs (Cite \cite{sakaguchi2000polarity-inversion-speech-perception})
Phase-intercept distortion manipulation will inevitably result in the start and end samples having different values, which may lead to audible artifacts and is not relevant to our hypothesis.
We thus applied a trapezoidal fade in/out to each stimuli, with a 0.1 second linear taper at both ends.
% Peak normalization was applied to both stimuli by the same amount across all questions, to ensure no clipping occurred, and no sudden changes in amplitude from one trial to the next.
To prevent clipping and ensure consistent amplitude between trials, we equally normalized the signal level for the pair of stimuli (distorted and unaltered) across all questions.

\subsection{Experiment Design}

\begin{figure}
    \centering
    \includegraphics[scale=0.26]{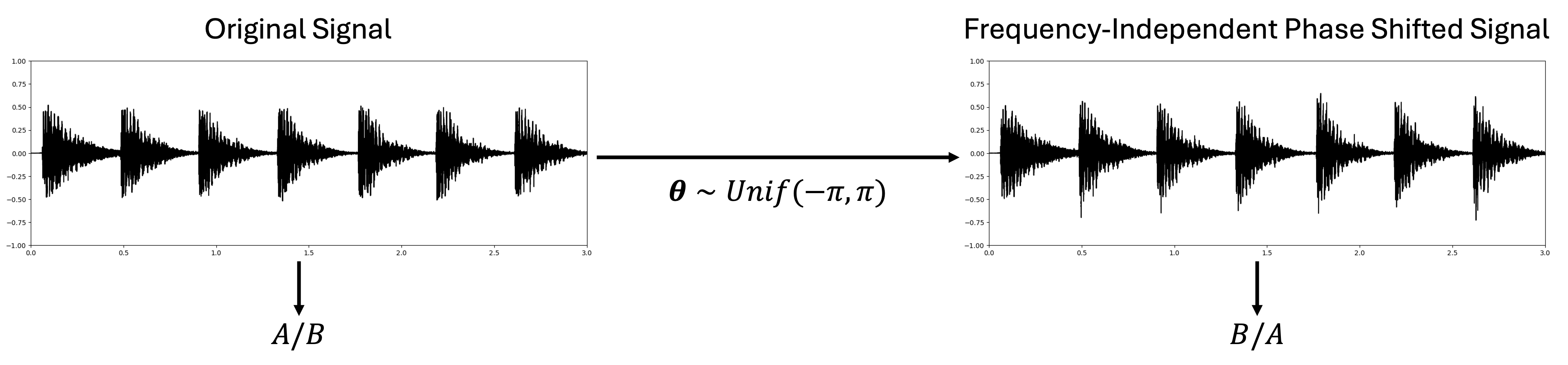}
    \caption{Illustration of the experimental design.}
    \label{fig:exp_design}
\end{figure}

%Assessing whether the differences between two stimuli (A and B) are perceptually relevant is a common goal in psychoacoustic research.
% Following standard practices, we designed a two-alternative forced choice (A/B) experiment, similar to the experiment conducted to measure perceptibility of phase distortion in all-pass filters \cite{deer1985perception-phase-distortion-all-pass}.
We employed a two-alternative forced-choice (A/B) design, similar to \cite{deer1985perception-phase-distortion-all-pass}.
% The benefit of this design is that if the subject is unable to hear a difference, the question still requires a response, forcing them to randomly guess between A and B.
% Thus, a result of 50\% accuracy would imply no perceptible difference on applying a frequency-independent phase shift.
The benefit of this design is that even if the subjects hear no difference, they must still guess, in which case their expected accuracy would be 50\%.
The experiment employed a within-subjects design, as all participants were exposed to all the stimuli; the order of stimuli was randomized for each participant.

\subsection{Procedure}

% Participants accessed the experiment via a web interface, created on Qualtrics, and were asked to use a good pair of headphones or a speaker system (such as studio monitors) to minimize information loss caused by poor audio equipment.
Participants accessed the experiment via a web interface, created on Qualtrics, and were instructed to use high-quality headphones or speakers (e.g., studio monitors) to minimize information loss caused by poor audio equipment.
Each participant completed thirty trials: ten music, ten speech, and ten other real-world sounds, with the total duration ranging from 10 to 15 minutes.

In each trial, participants were presented with reference stimuli and two comparison stimuli, labeled A and B; the original signal is randomly assigned as option A or B with 50\% probability.
The other option becomes the phase-intercept distorted version of the original signal.
The participant was then required to choose which signal (A or B) was ``identical'' to the original signal.
The experimental procedure can be summarized as shown in \cref{fig:exp_design}.
After the main experiment concluded, participants were asked to fill out a post-questionnaire to share their experience, in particular, any strategies they adopted during the main task.
Many participants commented that the pairs all sounded the same.

\section{Results}
\label{sec:results}

For each trial, the participant's response (A or B) can be coded correct or incorrect.
%We then observe the average accuracy of responses.
If our hypothesis is true, participants' true success rate should be approximately 50\%.
On average, our participants selected correctly in 49.87\% of trials (\cref{tab:phase_intercept_results}).
% A one sample $t$-test was performed to check if this accuracy is significantly different than the chance 50\% level was not significant.
A one-sample t-test was performed to check if this accuracy was significantly different from the chance level of 50\%, and the result was not significant.
% As seen in the overall results in \cref{tab:phase_intercept_results}, the resulting $t$-statistic is close to zero, with an associated $p$-value well above 0.7 (far from the typical 0.05 threshold required to reject the null hypothesis).
%The mean is close to 50\% with a small standard deviation, and the median is exactly 50\%, which also can be visualized using the box plot in .
However, a non-significant test result might arise from a small sample size, especially if the true effect is small.
We thus, focus on Bayesian statistical analysis, which is more appropriate for this analysis.

\begin{table}[t]
    \centering
    \caption{Perceptual Test Results (Percentage of Correct Answers)}
    \begin{tabular}{||c||c|c|c||c||}
        \hline
         & \textbf{Music} & \textbf{Speech} & \textbf{Other} & \textbf{Overall} \\
        \hline \hline
        Mean & 51.15\% & 45.77\% & 52.69\% & \textbf{49.87\%}\\
        \hline
        Median & 50\% & 50\% & 50\% & \textbf{50\%}\\
        \hline
        Standard deviation & 15.27\% & 19.05\% & 17.66\% & \textbf{9.36\%}\\
        \hline
        t-statistic & 0.378 & -1.111 & 0.762 & \textbf{-0.068}\\
        \hline
        p-value & 0.709 & 0.277 & 0.453 & \textbf{0.946}\\
        \hline
        Samples & \multicolumn{4}{c||}{26}\\
        \hline
    \end{tabular}
    \label{tab:phase_intercept_results}
\end{table}

\begin{figure}
    \centering
    \includegraphics[scale=0.22]{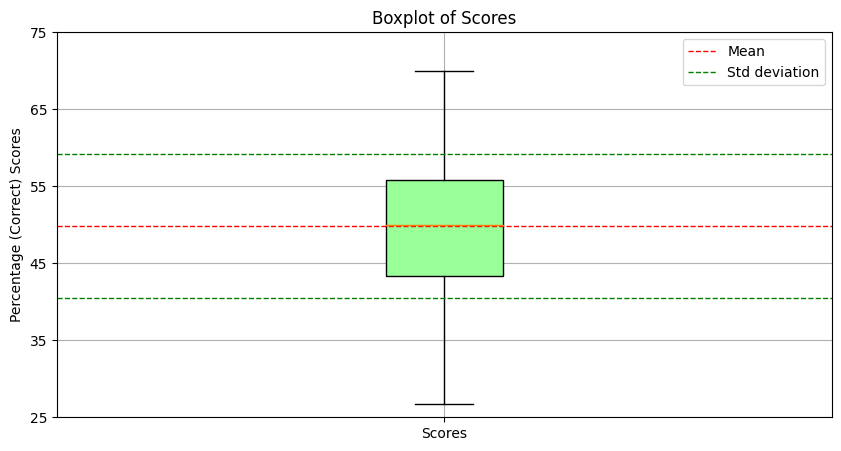}
    \caption{Distribution of response accuracy of 26 participants.}
    \label{fig:box_plot_intercept_distortion}
\end{figure}

Let $q$ be the true probability that a participant will pick the distorted recording from a pair of a stimuli.
Our hypothesis is that $q = 0.5$, corresponding to random binary guessing.
Consider a naive flat prior distribution for $q \sim \textnormal{Beta}(\alpha = 1,\ \beta = 1)$, corresponding to a neutral prior belief about the true value of $q$.
We can model each experimental trial as a single Bernoulli sample, where the outcome is either 0 (wrong pick) or 1 (correct pick), and the probability of the correct pick is $q$.
The Bernoulli distribution is the \emph{conjugate} distribution to the beta distribution (our prior distribution).
That means that the Bayesian posterior distribution after observing $N$ Bernoulli samples is also beta distributed, where $\alpha_{posterior} = \alpha_{prior} + N_{successes}$ and $\beta_{posterior} = \beta_{prior} + N_{failures}$.

\begin{figure}
    \centering
    \includegraphics[width=0.6\linewidth]{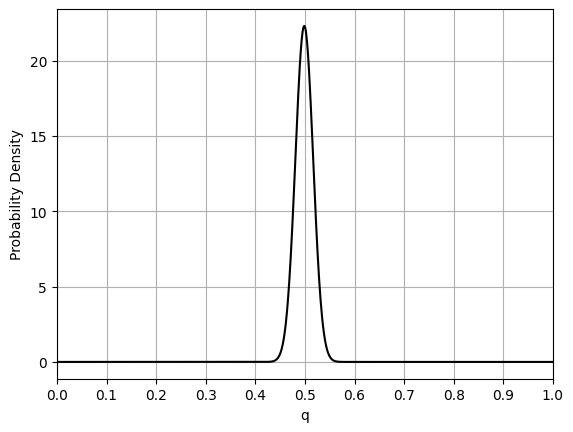}
    \caption{Posterior distribution of $q$, assuming flat prior distribution.}
    \label{fig:posteriorq}
\end{figure}

Out of the 780 total questions answered by the participants, they successfully selected the undistorted stimuli 389 times and failed 391 times.
Thus, the Bayesian posterior distribution for $q$, given the naive prior distribution we started with, is $q_{posterior} \sim \textnormal{Beta}(\alpha = 1 + 389,\ \beta = 1 + 391)$ (\cref{fig:posteriorq}).
As can be seen in \cref{fig:posteriorq}, even if we begin with flat prior belief about $q$, the data from this experiment leads to a posterior belief about $q$ that is tightly centered around $q = 0.5$, with 95\% of this posterior distribution falling between 0.464 and 0.534.
These results are consistent with the experimental hypothesis that phase-intercept distortion has no perceptible effects.

\cref{tab:phase_intercept_results} reports the results for each of the categories: music, speech and other.
The median results are 50\% for all categories.
The mean scores are similar for music and other, while the mean score of speech is worse than 50\%.
%This is most likely due to lack of samples.

% To test for the possibility of participants getting tired as the survey progressed, and should not be like they scored very well in the first few and then started going towards random behavior, we plotted the question-wise score and fit a line and analysed the slope.

\begin{figure}
    \centering
    \includegraphics[scale=0.21]{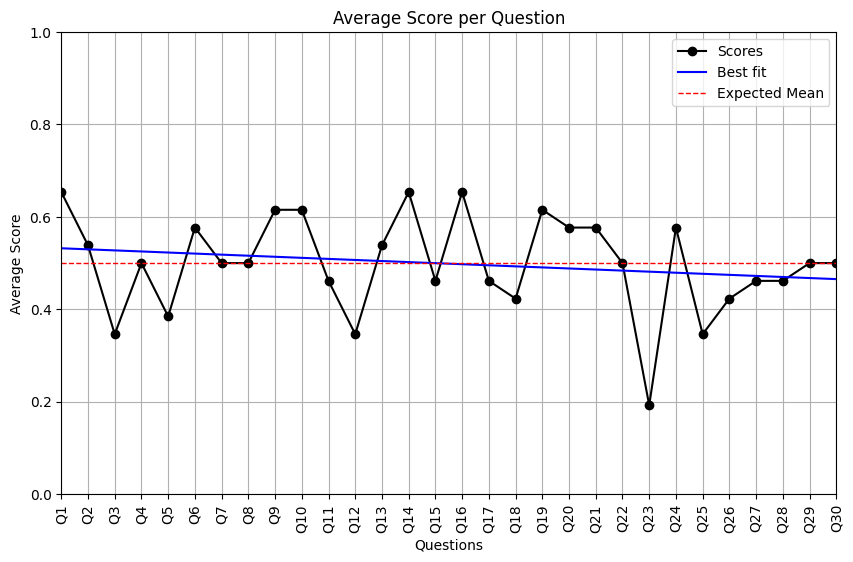}
    \caption{Question Scores}
    \label{fig:question_score_plot}
\end{figure}

% Participants' fatigue can affect performance over the course of the survey and result in performing well initially and gradually resorting to random responses.
% To ensure this was not the case, \cref{fig:question_score_plot} plots the question-wise scores, with a fitted trend line.
To ensure that participants' fatiguing over the course of the survey was not a confounding factor, \cref{fig:question_score_plot} plots the question-wise scores, with a fitted trend line.
This least-squares regression line has a negligible slope of $-0.002$.
This suggests that fatigue was not a confounding factor in our experiment.

% A non-significant test could simply arise if the sample size is too small, particularly if the true effect is small.
% For example, if people actually \emph{are} able to hear phase-intercept distortion slightly more often than chance---for example 55\% of the time---the statistical test might still be ``not significant'' because the sample size is too small to be sensitive to such a small effect.
% This means that the non-significant tests are difficult to interpret.
% A Bayesian statistical analysis can be better adapted to test our hypothesis.

% It would be good to look at the average accuracy of the different participants, and also to look at the average accuracy by audio type (music/speech/other). 

% (Should we plot how random chance would look like?)
% (Should we create a small accessible site where people can themselves test??)

\section{Data Augmentation Experiment}
\label{sec:data_augmentation}
% We shall discuss one of the applications of the perceptual invariance of phase-intercept distortion.
% We shall describe in detail one of the use cases where this concept can be applied -- \emph{data augmentation}.
% Datasets with limited amount of data can often lead to overfitting, which basically refers to a situation where a Machine Learning model learns the training data too much precisely that it fails to generalize to different situations.
% This is often seen with very good training results, accompanied by a poor validation and test results.
In Machine Learning (ML), data augmentation is the process of performing operations to increase the amount of input data to achieve better results \cite{watcharasupat2024stem, gong2021ast, yu2021frequency, manilow2022improvingBSS}.
It helps to counter overfitting, a common issue in ML where the model learns the training data too precisely and fails to generalize.

Commonly used data augmentation techniques for audio ML tasks include pitch shifting, time stretching, speed perturbation, noise addition, harmonic distortion, random time-frequency masking, and amplitude gain \cite{nanni2020audio_data_aug, ko2015speech_augmentation, wei2020comparison_data_augmentation}.
The effect of image-based data augmentation styles on magnitude spectrograms have also been researched \cite{nanni2020audio_data_aug}.
These strategies have improved results for various audio-based tasks.

We propose using \textit{phase-intercept distortion} as a data augmentation technique for music and speech ML tasks, such as classification, source separation, and generation.
With an overall computational complexity of $\mathcal{O}(n \log n)$, this form of data augmentation is a fast and unique way to modify a signal without altering the transients, the pitch, and time characteristics of the signal.
Data augmentation can be performed by randomly sampling a $\theta \in [-\pi, \pi)$ and modifying the signal using \cref{eq:freq_ind_phase_shift}.
This can be considered as a generalized version of the more restrictive Inverse Phase Augmentation (IPA) \cite{catellier2020wawenets, catellier2023wideband}.
Since phase-intercept distortion does not affect the magnitude spectrogram (as shown in \cref{eq:freq-response-phase-intercept-distortion}), this augmentation is not applicable to models using magnitude spectrograms as inputs---it is only applicable to those using time-domain signals or complex spectrograms.

To truly test whether the benefits of data augmentation stem from the diversity it provides and not the quantity of data, we designed our experiments with data augmentation added ``on the fly'' during training.
This means that each time a training sample is input to the model, it is randomly transformed in real time.
This ensures variation without increasing the dataset size, and isolates the impact of diversity provided by data augmentation.

We performed experiments on two tasks: audio classification and source separation.
We used models that take time-domain audio signals as input for these tasks.

\subsection{Audio Classification}
Audio classification is the task of assigning audio signals to corresponding classes based on their characteristics.
This process typically involves an initial stage of feature extraction from the raw audio.
% CITE!
% Some typical features which are extracted include (i) spectrograms, which are generated using the Short-Time Fourier Transform (STFT) to represent the signal’s frequency content over time, (ii) mel spectrograms, which are calculated by applying the mel-scale filterbank on a spectrogram to better align with human auditory perception, and (iii) mel frequency cepstral coefficients or MFCCs, which are used for capturing timbre information of the audio.
Some typical features include spectrograms, mel-spectrograms, and mel-frequency cepstral coefficients or MFCCs.
These classification models then learn to map these features to the audio classes.

We chose to use the SC09 dataset for the task, a subset of the Speech Commands dataset \cite{warden2018speech-commands}, which includes spoken digits from zero to nine from various different speakers.
% Models used before to train on these to get results

Choosing models to test was more difficult, because few existing audio classification models that use time-domain signals as a direct input.
We thus designed a new model for this experiment, taking inspiration from wav2vec2.0 \cite{baevski2020wav2vec2.0}.
We chose the exact same convolution layers as the initial layers of wav2vec2.0 to encode the audio information, and used a two layer bidirectional LSTM \cite{graves2005bilstm} on these encodings.
Batch normalization \cite{ioffe2015batchnorm} was performed in between each layer with a dropout probability of 0.05.
Finally, the output from either side was concatenated and passed through a fully connected layer for the prediction.
% The overall architecture can be seen in \cref{}.

Five different seed values were chosen randomly for the experiment.
The mean and standard deviation of the validation and test accuracies are reported in \cref{tab:classification}.
The validation accuracy shows improvement when phase-intercept distortion is used as a data augmentation technique for this model.
However, the test accuracy is only slightly better.

\begin{table}[t]
\centering
\caption{Phase intercept distortion as data augmentation for classification}
\label{tab:classification}
\sisetup{
    reset-text-series = false, 
    text-series-to-math = true, 
    mode=text,
    tight-spacing=true,
    round-mode=places,
    round-precision=2,
    table-format=2.2,
    table-number-alignment=center
}
\begin{tabular}{l*{1}{S[round-precision=1,table-format=2.1]S}}
    \toprule
    & {Validation Accuracy $\uparrow$}& {Test Accuracy $\uparrow$}\\
    \midrule
    No Augmentation &            \text{93.56 ± 0.30} \% &           \text{92.41 ± 0.40} \% \\
    \midrule
    With Augmentation &            \textbf{93.99 ± 0.26} \% &         \textbf{92.43 ± 0.35} \% \\
    \bottomrule
\end{tabular}
\end{table}

\subsection{Blind Source Separation}
Blind source separation (BSS) is defined as the task of separating the individual sources when only provided with a single mixture audio signal.
Supervised approaches have been the go-to strategy due to their strong performance, which is achieved by learning the mapping between mixtures and separated components.
There are three primary ways to solve this: (i) waveform prediction \cite{stoller2018waveUnet, plaja2022diffusion-waveform-source-separation, defossez2019demucs}, (ii) spectrogram prediction \cite{hennequin2020spleeter, jansson2017SVS-separation-spec-based}, and (iii) hybrid methods, which combine elements of both (i) and (ii) \cite{defossez2021hybrid-demucs, rouard2023htdemucs}.

\begin{table}[t]
\centering
\caption{Phase intercept distortion as data augmentation for source separation}
\label{tab:bss_expt}
\sisetup{
    reset-text-series = false, 
    text-series-to-math = true, 
    mode=text,
    tight-spacing=true,
    round-mode=places,
    round-precision=2,
    table-format=2.2,
    table-number-alignment=center
}
\begin{tabular}{l*{3}{S[round-precision=1,table-format=3.2]S}}
    \toprule
    & {SDR $\uparrow$} & {SI-SDR $\uparrow$} & {SIR $\uparrow$} & {SAR $\uparrow$} & {ISR $\uparrow$}\\
    \midrule
    No Augmentation  & \text{3.58} &  \text{1.27} & \text{4.99} & \textbf{2.76} & \text{5.94}\\
    \midrule
    With Augmentation & \textbf{3.76} & \textbf{1.71} & \textbf{5.21} & \text{2.76} & \textbf{5.98} \\
    \bottomrule
\end{tabular}
\end{table}

We chose a popular waveform prediction-based model called the Wave-U-Net \cite{stoller2018waveUnet}, which works directly on source image waveforms.
The details of the metrics are described in \cite{vincent2007sisec2007}.
\cref{tab:bss_expt} reports the median results of this experiment in decibels (dB).
SIR, SAR and ISR are computed using the \texttt{bss\_eval} toolkit \cite{vincent2006bssevaltoolbox}.
For the individual examples we compute the median value of all the frames.
Signal to Distortion Ratio (SDR) and Scale Invariant Signal to Distortion Ratio (SI-SDR) are calculated as described in \cite{le2019sisdr}.
Phase intercept distortion as a data augmentation technique for Wave-U-Net on MUSDB18-HQ has shown improvement in all the metrics, except SAR.

% \begin{table}[t]
% \centering
% \caption{Effect of phase-intercept distortion as data augmentation on a source separation task}
% \label{tab:bss_expt}
% \sisetup{
%     reset-text-series = false, 
%     text-series-to-math = true, 
%     mode=text,
%     tight-spacing=true,
%     round-mode=places,
%     round-precision=2,
%     table-format=2.2,
%     table-number-alignment=center
% }
% \begin{tabular}{l*{2}{S[round-precision=1,table-format=3.2]S}}
%     \toprule
%     & {SNR $\uparrow$} & {SI-SNR $\uparrow$} & {SIR $\uparrow$}\\
%     \midrule
%     No Augmentation  & \text{3.55 ± 2.22} &  \text{0.66 ± 4.91} & \text{3.11 ± 9.72} \\
%     \midrule
%     Added Augmentation & \textbf{3.72 ± 2.41} & \textbf{0.88 ± 5.22} & \textbf{3.87 ± 9.01} \\
%     \bottomrule
% \end{tabular}
% \begin{tabular}{l*{1}{S[round-precision=1,table-format=3.2]S}}
%     \toprule
%     & {SAR $\uparrow$} & {ISR $\uparrow$}\\
%     \midrule
%     No Augmentation  &  \textbf{3.07 ± 2.76} & \text{4.61 ± 4.41} \\
%     \midrule
%     Added Augmentation &  \text{2.93 ± 2.87} & \textbf{5.11 ± 4.38}  \\
%     \bottomrule
% \end{tabular}
% \end{table}
\section{Conclusion and Future Work}
This research study showed that monaural phase-intercept distortion is perceptually invariant for general sounds humans hear in their daily lives.
However, this result can only be used for signals that humans can hear, and will be incorrect to use in biomedical signal analysis.
%This is also not the solution for perceptual invariance but a step towards it.
Our initial experiments here hint that data augmentation through phase-intercept distortion may indeed be a viable approach to improving ML models for classification and source separation.
However, more experiments with diverse datasets and models are required to confirm that models consistently benefit from this form of data augmentation.

The monaural perceptual invariance of phase-intercept distortion for general sounds opens up various research possibilities beyond data augmentation, including the improvement of evaluation metrics for various audio ML tasks.
While these directions remain outside the scope of this work, they offer promising avenues for future exploration.
Leveraging this perceptual invariance for applications where frequency independent phase shifting can potentially cause problems, will be a step towards more robust and perceptually aligned models.

\clearpage
% The \IEEEtriggeratref{XX} command can be used to move to the next column before the XX-th reference
% to balance the two columns of the reference section
% \IEEEtriggeratref{XX}
\bibliographystyle{IEEEtran}
\bibliography{WASPAA2025_paper_template}

\begin{thebibliography}{10}
\providecommand{\url}[1]{#1}
\csname url@samestyle\endcsname
\providecommand{\newblock}{\relax}
\providecommand{\bibinfo}[2]{#2}
\providecommand{\BIBentrySTDinterwordspacing}{\spaceskip=0pt\relax}
\providecommand{\BIBentryALTinterwordstretchfactor}{4}
\providecommand{\BIBentryALTinterwordspacing}{\spaceskip=\fontdimen2\font plus
\BIBentryALTinterwordstretchfactor\fontdimen3\font minus \fontdimen4\font\relax}
\providecommand{\BIBforeignlanguage}[2]{{%
\expandafter\ifx\csname l@#1\endcsname\relax
\typeout{** WARNING: IEEEtran.bst: No hyphenation pattern has been}%
\typeout{** loaded for the language `#1'. Using the pattern for}%
\typeout{** the default language instead.}%
\else
\language=\csname l@#1\endcsname
\fi
#2}}
\providecommand{\BIBdecl}{\relax}
\BIBdecl

\bibitem{ohm1843definition}
G.~S. Ohm, ``{\"U}ber die definition des tones, nebst daran gekn{\"u}pfter theorie der sirene und {\"a}hnlicher tonbildender vorrichtungen,'' \emph{Annalen der Physik}, vol. 135, no.~8, pp. 513--565, 1843.

\bibitem{nyquist1930measurement-phase-distortion}
H.~Nyquist and S.~Brand, ``Measurement of phase distortion,'' \emph{Bell System Technical Journal}, vol.~9, no.~3, pp. 522--549, 1930.

\bibitem{preis1982phase-distortion-tutorial}
D.~Preis, ``Phase distortion and phase equalization in audio signal processing-a tutorial review,'' \emph{Journal of the Audio Engineering Society}, vol.~30, no.~11, pp. 774--794, 1982.

\bibitem{plomp1969perception_complex_tone_phase}
R.~Plomp and H.~J. Steeneken, ``Effect of phase on the timbre of complex tones,'' \emph{The Journal of the Acoustical Society of America}, vol.~46, no.~2B, pp. 409--421, 1969.

\bibitem{deer1985perception-phase-distortion-all-pass}
J.~Deer, P.~Bloom, and D.~Preis, ``Perception of phase distortion in all-pass filters,'' \emph{Journal of the Audio Engineering Society}, vol.~33, no.~10, pp. 782--786, 1985.

\bibitem{preis1983perception-phase-distortion-anti-alias}
D.~Preis and P.~Bloom, ``Perception of phase distortion in anti-alias filters,'' in \emph{Audio Engineering Society Convention 74}.\hskip 1em plus 0.5em minus 0.4em\relax Audio Engineering Society, 1983.

\bibitem{lipshitz1982audibility-phase-intercept-distortion}
S.~P. Lipshitz, M.~Pocock, and J.~Vanderkooy, ``On the audibility of midrange phase distortion in audio systems,'' \emph{Journal of the Audio Engineering Society}, vol.~30, no.~9, pp. 580--595, 1982.

\bibitem{craig1962perception_two_tone_phase}
J.~H. Craig and L.~A. Jeffress, ``Effect of phase on the quality of a two-component tone,'' \emph{The Journal of the Acoustical Society of America}, vol.~34, no.~11, pp. 1752--1760, 1962.

\bibitem{hansen1974aural}
V.~Hansen and E.~R. Madsen, ``On aural phase detection: Part 1,'' \emph{J. Audio Eng. Soc}, vol.~22, no.~1, pp. 10--14, 1974.

\bibitem{von1912helmholtz}
H.~Von~Helmholtz, \emph{On the Sensations of Tone as a Physiological Basis for the Theory of Music}.\hskip 1em plus 0.5em minus 0.4em\relax Longmans, Green, 1912.

\bibitem{bregman1994auditory-scene-analysis}
A.~S. Bregman, \emph{Auditory scene analysis: The perceptual organization of sound}.\hskip 1em plus 0.5em minus 0.4em\relax MIT press, 1994.

\bibitem{suzuki1980perception-phase-distortion}
H.~Suzuki, S.~Morita, and T.~Shindo, ``On the perception of phase distortion,'' \emph{Journal of the Audio Engineering Society}, vol.~28, no.~9, pp. 570--574, 1980.

\bibitem{chappel2016phase-distortion-speech}
R.~Chappel, B.~Schwerin, and K.~Paliwal, ``Phase distortion resulting in a just noticeable difference in the perceived quality of speech,'' \emph{Speech Communication}, vol.~81, pp. 138--147, 2016.

\bibitem{bedrosian2007analyticrespresentation}
E.~Bedrosian, ``The analytic signal representation of modulated waveforms,'' \emph{Proceedings of the IRE}, vol.~50, no.~10, pp. 2071--2076, 2007.

\bibitem{gabor1946theoryofcommunication}
D.~Gabor, ``Theory of communication. part 1: The analysis of information,'' \emph{Journal of the Institution of Electrical Engineers-part III: radio and communication engineering}, vol.~93, no.~26, pp. 429--441, 1946.

\bibitem{yang2017analyticsignalenvelope}
Y.~Yang, ``A signal theoretic approach for envelope analysis of real-valued signals,'' \emph{IEEE Access}, vol.~5, pp. 5623--5630, 2017.

\bibitem{taylor1981phase}
L.~Taylor, ``The phase retrieval problem,'' \emph{IEEE Transactions on Antennas and Propagation}, vol.~29, no.~2, pp. 386--391, 1981.

\bibitem{huang2009instantaneousFrequencyAnalytic}
N.~E. Huang, Z.~Wu, S.~R. Long, K.~C. Arnold, X.~Chen, and K.~Blank, ``On instantaneous frequency,'' \emph{Advances in adaptive data analysis}, vol.~1, no.~2, pp. 177--229, 2009.

\bibitem{sakaguchi2000polarity-inversion-speech-perception}
S.~Sakaguchi, T.~Arai, and Y.~Murahara, ``The effect of polarity inversion of speech on human perception and data hiding as an application,'' in \emph{2000 IEEE International Conference on Acoustics, Speech, and Signal Processing. Proceedings (Cat. No.00CH37100)}, vol.~2, 2000, pp. II917--II920 vol.2.

\bibitem{gemmeke2017audioset}
J.~F. Gemmeke, D.~P. Ellis, D.~Freedman, A.~Jansen, W.~Lawrence, R.~C. Moore, M.~Plakal, and M.~Ritter, ``Audio set: An ontology and human-labeled dataset for audio events,'' in \emph{2017 IEEE international conference on acoustics, speech and signal processing (ICASSP)}.\hskip 1em plus 0.5em minus 0.4em\relax IEEE, 2017, pp. 776--780.

\bibitem{rafii2019musdb18hq}
\BIBentryALTinterwordspacing
Z.~Rafii, A.~Liutkus, F.-R. St{\"o}ter, S.~I. Mimilakis, and R.~Bittner, ``{MUSDB18-HQ} - an uncompressed version of musdb18,'' Dec. 2019. [Online]. Available: \url{https://doi.org/10.5281/zenodo.3338373}
\BIBentrySTDinterwordspacing

\bibitem{mathes1947perception_mpe}
R.~Mathes and R.~Miller, ``Phase effects in monaural perception,'' \emph{The Journal of the Acoustical Society of America}, vol.~19, no.~5, pp. 780--797, 1947.

\bibitem{srinivasamurthy2021saraga}
A.~Srinivasamurthy, S.~Gulati, R.~C. Repetto, and X.~Serra, ``Saraga: Open datasets for research on indian art music,'' \emph{Empirical Musicology Review}, vol.~16, no.~1, pp. 85--98, 2021.

\bibitem{krishnan2025sanidha}
V.~V. Krishnan, N.~Alben, A.~Nair, and N.~Condit-Schultz, ``Sanidha: A studio quality multi-modal dataset for carnatic music,'' \emph{arXiv preprint arXiv:2501.06959}, 2025.

\bibitem{panayotov2015librispeech}
V.~Panayotov, G.~Chen, D.~Povey, and S.~Khudanpur, ``Librispeech: an asr corpus based on public domain audio books,'' in \emph{2015 IEEE international conference on acoustics, speech and signal processing (ICASSP)}.\hskip 1em plus 0.5em minus 0.4em\relax IEEE, 2015, pp. 5206--5210.

\bibitem{watcharasupat2024stem}
K.~N. Watcharasupat and A.~Lerch, ``A stem-agnostic single-decoder system for music source separation beyond four stems,'' \emph{arXiv preprint arXiv:2406.18747}, 2024.

\bibitem{gong2021ast}
Y.~Gong, Y.-A. Chung, and J.~Glass, ``Ast: Audio spectrogram transformer,'' \emph{arXiv preprint arXiv:2104.01778}, 2021.

\bibitem{yu2021frequency}
S.~Yu, X.~Sun, Y.~Yu, and W.~Li, ``Frequency-temporal attention network for singing melody extraction,'' in \emph{ICASSP 2021-2021 IEEE International Conference on Acoustics, Speech and Signal Processing (ICASSP)}.\hskip 1em plus 0.5em minus 0.4em\relax IEEE, 2021, pp. 251--255.

\bibitem{manilow2022improvingBSS}
E.~Manilow, C.~Hawthorne, C.-Z.~A. Huang, B.~Pardo, and J.~Engel, ``Improving source separation by explicitly modeling dependencies between sources,'' in \emph{ICASSP 2022-2022 IEEE International Conference on Acoustics, Speech and Signal Processing (ICASSP)}.\hskip 1em plus 0.5em minus 0.4em\relax IEEE, 2022, pp. 291--295.

\bibitem{nanni2020audio_data_aug}
L.~Nanni, G.~Maguolo, and M.~Paci, ``Data augmentation approaches for improving animal audio classification,'' \emph{Ecological Informatics}, vol.~57, p. 101084, 2020.

\bibitem{ko2015speech_augmentation}
T.~Ko, V.~Peddinti, D.~Povey, and S.~Khudanpur, ``Audio augmentation for speech recognition.'' in \emph{Interspeech}, vol. 2015, 2015, p. 3586.

\bibitem{wei2020comparison_data_augmentation}
S.~Wei, S.~Zou, F.~Liao \emph{et~al.}, ``A comparison on data augmentation methods based on deep learning for audio classification,'' in \emph{Journal of physics: Conference series}, vol. 1453, no.~1.\hskip 1em plus 0.5em minus 0.4em\relax IOP Publishing, 2020, p. 012085.

\bibitem{catellier2020wawenets}
A.~A. Catellier and S.~D. Voran, ``Wawenets: A no-reference convolutional waveform-based approach to estimating narrowband and wideband speech quality,'' in \emph{ICASSP 2020-2020 IEEE International Conference on Acoustics, Speech and Signal Processing (ICASSP)}.\hskip 1em plus 0.5em minus 0.4em\relax IEEE, 2020, pp. 331--335.

\bibitem{catellier2023wideband}
------, ``Wideband audio waveform evaluation networks: Efficient, accurate estimation of speech qualities,'' \emph{IEEE Access}, vol.~11, pp. 125\,576--125\,592, 2023.

\bibitem{warden2018speech-commands}
P.~Warden, ``Speech commands: A dataset for limited-vocabulary speech recognition,'' \emph{arXiv preprint arXiv:1804.03209}, 2018.

\bibitem{baevski2020wav2vec2.0}
A.~Baevski, Y.~Zhou, A.~Mohamed, and M.~Auli, ``Wav2vec 2.0: A framework for self-supervised learning of speech representations,'' \emph{Advances in neural information processing systems}, vol.~33, pp. 12\,449--12\,460, 2020.

\bibitem{graves2005bilstm}
A.~Graves and J.~Schmidhuber, ``Framewise phoneme classification with bidirectional lstm networks,'' in \emph{Proceedings. 2005 IEEE International Joint Conference on Neural Networks, 2005.}, vol.~4.\hskip 1em plus 0.5em minus 0.4em\relax IEEE, 2005, pp. 2047--2052.

\bibitem{ioffe2015batchnorm}
S.~Ioffe and C.~Szegedy, ``Batch normalization: Accelerating deep network training by reducing internal covariate shift,'' in \emph{International conference on machine learning}.\hskip 1em plus 0.5em minus 0.4em\relax pmlr, 2015, pp. 448--456.

\bibitem{stoller2018waveUnet}
D.~Stoller, S.~Ewert, and S.~Dixon, ``Wave-u-net: A multi-scale neural network for end-to-end audio source separation,'' \emph{arXiv preprint arXiv:1806.03185}, 2018.

\bibitem{plaja2022diffusion-waveform-source-separation}
G.~Plaja-Roglans, M.~Miron, and X.~Serra, ``A diffusion-inspired training strategy for singing voice extraction in the waveform domain,'' 2022.

\bibitem{defossez2019demucs}
A.~D{\'e}fossez, N.~Usunier, L.~Bottou, and F.~Bach, ``Music source separation in the waveform domain,'' \emph{arXiv preprint arXiv:1911.13254}, 2019.

\bibitem{hennequin2020spleeter}
R.~Hennequin, A.~Khlif, F.~Voituret, and M.~Moussallam, ``Spleeter: a fast and efficient music source separation tool with pre-trained models,'' \emph{Journal of Open Source Software}, vol.~5, no.~50, p. 2154, 2020.

\bibitem{jansson2017SVS-separation-spec-based}
A.~Jansson, E.~Humphrey, N.~Montecchio, R.~Bittner, A.~Kumar, and T.~Weyde, ``Singing voice separation with deep u-net convolutional networks,'' 2017.

\bibitem{defossez2021hybrid-demucs}
A.~D{\'e}fossez, ``Hybrid spectrogram and waveform source separation,'' \emph{arXiv preprint arXiv:2111.03600}, 2021.

\bibitem{rouard2023htdemucs}
S.~Rouard, F.~Massa, and A.~D{\'e}fossez, ``Hybrid transformers for music source separation,'' in \emph{ICASSP 2023-2023 IEEE International Conference on Acoustics, Speech and Signal Processing (ICASSP)}.\hskip 1em plus 0.5em minus 0.4em\relax IEEE, 2023, pp. 1--5.

\bibitem{vincent2007sisec2007}
E.~Vincent, H.~Sawada, P.~Bofill, S.~Makino, and J.~P. Rosca, ``First stereo audio source separation evaluation campaign: data, algorithms and results,'' in \emph{International Conference on Independent Component Analysis and Signal Separation}.\hskip 1em plus 0.5em minus 0.4em\relax Springer, 2007, pp. 552--559.

\bibitem{vincent2006bssevaltoolbox}
E.~Vincent, R.~Gribonval, and C.~F{\'e}votte, ``Performance measurement in blind audio source separation,'' \emph{IEEE transactions on audio, speech, and language processing}, vol.~14, no.~4, pp. 1462--1469, 2006.

\bibitem{le2019sisdr}
J.~Le~Roux, S.~Wisdom, H.~Erdogan, and J.~R. Hershey, ``{SDR}--half-baked or well done?'' in \emph{ICASSP 2019-2019 IEEE International Conference on Acoustics, Speech and Signal Processing (ICASSP)}.\hskip 1em plus 0.5em minus 0.4em\relax IEEE, 2019, pp. 626--630.

\end{thebibliography}
% or list them by yourself:

% \bibitem{waspaaweb}
% {WASPAA Website}, \url{http://www.waspaa.com}.

% \bibitem{IEEEXploreReqs}
% {IEEE {X}plore {R}equirements}, \url{https://conferences.ieeeauthorcenter.ieee.org/write-your-paper/meet-ieee-xplore-requirements/}.

% \bibitem{eWilliams1999}
% E.~Williams, \emph{Fourier Acoustics: Sound Radiation and Nearfield Acoustic Holography}.\hskip 1em plus 0.5em minus 0.4em\relax London, UK: Academic Press, 1999.

% \bibitem{cJones2003}
% C.~Jones, A.~Smith, and E.~Roberts, ``A sample paper in conference proceedings,'' in \emph{Proc. ICASSP}, vol.~II, Apr. 2003, pp. 803--806.

% \bibitem{aSmith2000}
% A.~Smith, C.~Jones, and E.~Roberts, ``A sample paper in journals,'' \emph{IEEE Trans. Signal Process.}, vol.~62, pp. 291--294, Jan. 2000.

% \end{thebibliography}

\end{document}